# Atomic Gold and Palladium Anion-Catalysis of Water to Peroxide: Fundamental Mechanism

**Aron Tesfamichael[1], Kelvin Suggs[1], Zineb Felfli[2*], Xiao-Qian Wang[2] and Alfred Z. Msezane[2]**

[1]Department of Chemistry, Clark Atlanta University, Atlanta, Georgia 30314, USA
[2]Department of Physics and Center for Theoretical Studies of Physical Systems, Clark Atlanta University, Atlanta, Georgia 30314, USA

(*) E-mail: zfelfli@cau.edu

## ABSTRACT
We have performed dispersion-corrected density-functional transition state calculations on atomic $Au^-$ and $Pd^-$ catalysis of water conversion to peroxide. The $Au^-$ ion is found to be an excellent catalyst; however, atomic $Pd^-$ has a higher catalytic effect on the formation of peroxide. The $Au^-(H_2O)_2$ and $Pd^-(H_2O)_2$ anion molecular complexes formation in the transition state is identified as the fundamental mechanism for breaking the hydrogen bonding strength during the water catalysis. Our theoretical results provide crucial insight into the mechanism of the atomic $Au^-$ and $Pd^-$ catalysis in good agreement with recent experimental observations.

## 1. Introduction

The role played by atomic particles and nanoparticles in catalysis has attracted a wide range of fundamental and industrial interests in recent years [1-16]. Recent advances have demonstrated potential applications of gold nanoparticle catalysts in fuel cell related catalytic reactions, including in the catalytic partial oxidation of methane into valuable by-products, which is of great scientific importance and considerable industrial, economic and environmental interest. Furthermore, the catalytic activity of the Au–Ag–Pd trimetallic nanoparticles was found to be efficient in catalyzing the Heck reaction [14]. The experimental investigation [15] found that the methanol oxidation current of the ternary Pt–Ru–Ni catalyst increased significantly in comparison with that of the binary Pt–Ru catalyst. When the above referenced papers are examined carefully, it is clear that more effective nanocatalysts consist of at least two different atoms. From the configuration of the negative-ion resonances and Ramsauer–Townsend (R-T) minima in the low-energy electron elastic total cross sections (TCSs) of the atoms Au, Y, Ru, Pd, Ag and Pt, it was concluded that these atoms represent excellent possible candidates individually or in various combinations for nanocatalysts [17]. This gives the possibility to systematically construct effective nanocatalysts from these atoms for various substances by assembling them together atom by atom in various combinations. A recent interesting and revealing study has pointed out that the characteristic minimum in the photoionization cross section has the same origin as the R-T effect in low-energy elastic scattering [18]; the connection of the R-T minimum in very low-energy $F+H_2$ elastic scattering with a virtual state formation has been discussed theoretically [19].

Recently, the direct synthesis of hydrogen peroxide from $H_2$ and $O_2$ using supported Au, Pd and Au-Pd nanoparticle catalysts was reported [9,16]. The experimental studies revealed that the addition of Pd to the Au nanocatalyst increased the rate of $H_2O_2$ synthesis significantly as well as the concentration of the $H_2O_2$ formed. Lacking in these investigations is a fundamental understanding at the atomic and/or molecular level how both the Au and Pd nanoparticles catalyze the reaction. Consequently, in this work we have carried out quantum calculations to elucidate the mechanism underlying the Au and Pd nanoparticles' excellent catalytic properties [9,16], including the substantial enhancement of the Au-Pd nanocatalyst over the individual Pd and Au. Specifically, we investigate the transition state of the oxidation of water to peroxide by performing dispersion-corrected density-functional theory calculations on the catalytic properties of atomic $Au^-$ and atomic $Pd^-$. The main objective is to complete our fundamental understanding of nanocatalysis of $H_2O_2$ from $H_2O$ from the chemical reaction dynamics perspective.

The near-threshold electron elastic scattering total cross sections for both the ground and excited states of simple and complex atoms have been found to be characterized by Ramsauer-Townsend minima, shape resonances, and extremely sharp resonances. These dramatically sharp long-lived resonances have been identified with the formation of stable bound states of the relevant negative ions formed during the elastic collision between the incident electron and the target neutral atom as Regge resonances [20, 21]. The recently developed Regge-pole, also known as the complex angular momentum (CAM) methodology wherein the crucial electron-electron correlations are embedded has been employed for the calculations. The vital core polarization interaction is incorporated through the well-investigated Thomas-Fermi type potential. The great strength and advantages of the CAM method over many existing sophisticated theoretical approaches, particularly the structure-based methods, to calculate accurate low-energy electron scattering resonances, including the attendant extraction of reliable electron affinities of simple and complex atoms, has been carefully discussed in [21]. Reliable atomic and molecular affinities, manifesting the existence of long-lived negative ion formation, are crucial for understanding a large number of chemical reactions involving negative ions [22]. The role of resonances is to promote anion formation through electron attachment [23]. These references demonstrate the importance and the need of identifying and delineating the resonances in low-energy electron scattering.

The same fundamental mechanism that underlies the well-investigated muon-catalyzed nuclear fusion using deuterium and tritium has been proposed to drive nanoscale catalysis [24, 25], see also [26] and references therein. Specifically, the fundamental atomic mechanism responsible for the oxidation of water to peroxide has been attributed to the interplay between Regge resonances and R-T minima in the electron elastic TCSs for the Au and Pd atoms, along with their large electron affinities (EAs) [24].

## 2. Method of calculation

We have employed first principles calculations based on Density Functional Theory (DFT) and dispersion-corrected DFT approaches to investigate atomic $Au^-$ and $Pd^-$ particle catalysis of water conversion to peroxide. Local-density-approximation (LDA) with Vosko, Wilk, and Nusair (VWN) functional was used for the preliminary geometry optimization. For preselecting structural molecular confirmation we further utilized the gradient-corrected Perdew-Burke-Ernzerfof(PBE) parameterizations[27] of the exchange-correlation rectified with the dispersion corrections [28]. The double numerical plus polarization basis set was employed as implemented in the DMol$_3$ package [29]. The dispersion-correction method, coupled to suitable density functional, has been demonstrated to account for the long-range dispersion forces with remarkable accuracy. We used a tolerance of $1.0 \times 10^{-3}$ eV for the energy convergence. A transition-state search employing nudged elastic bands facilitates the evaluation of energy barriers [30,31,32].

## 3.  Results and Discussion

We first consider the slow oxidation of water to peroxide without the presence of a catalyst:

$$2H_2O + O_2 \rightarrow 2H_2O_2 \qquad (1)$$

Then we apply the atomic Au$^-$ ion to speed up the reaction (1) and obtain

$$Au^-(H_2O)_2 + O_2 \rightarrow Au^- + 2\,H_2O_2. \qquad (2)$$

$$Au^- + 4H_2O + O_2 \rightarrow Au^-(H_2O)_2 + 2\,H_2O_2. \qquad (3)$$

Add reactions (2) and (3) and obtain

$$4H_2O + 2O_2 \rightarrow 4H_2O_2 \qquad (4)$$

A similar result as in (4) is obtained when the Pd$^-$ ion is substituted for Au$^-$. The anionic complexes Au$^-$(H$_2$O)$_2$ and Pd$^-$(H$_2$O)$_2$ have been characterized as atomic Au$^-$ and Pd$^-$ ions interacting with two water molecules, respectively, i.e., as the anion-molecule complexes[33]. The large electron affinities of the Au and Pd atoms played essential roles; they are important in the dissociation of the Au$^-$(H$_2$O)$_2$ and Pd$^-$(H$_2$O)$_2$ complexes breaking into Au$^-$ or Pd$^-$ and (H$_2$O)$_2$, respectively[33]. A recent experiment [11] determined the vertical detachment energies (VDEs) of the Au$^-$M complexes (M = Ne, Ar, Kr, Xe, O$_2$, CH$_4$ and H$_2$O). Importantly, the VDEs are all within the location in the energy of the second R–T (absolute) minimum in the TCS of atomic Au, which is also where Au has its EA. Furthermore, the experiment found a stronger interaction between the Au$^-$ and H$_2$O and that Au$^-$ does not react with O$_2$. In essence, these two experiments [11, 33] probed nanogold catalysis at its most fundamental level. Their results are important in the fundamental understanding of the extensive catalytic property of nanogold.

The breaking of hydrogen bonding is attributed to the formation of nanocatalyst anionic complexes involving an Au$^-$ ion and two water molecules. Water possesses the unique properties that are rare in other materials and are of biological importance. These properties are evident in hydrogen bonded environments, particularly in liquid water. In liquid water, the hydrogen bond's enthalpy is approximately 5.5 kcal/mol, and the total dissociation energy is 117 kcal/mol. Hydrogen bonding has a direct effect on the change in the Gibbs free energy, $G$ ($\Delta G = \Delta H - T\Delta S$), where $H$, $T$ and $S$ represent enthalpy, temperature, and entropy, respectively. The introduction of atomic Au$^-$ and/or Pd$^-$ into the oxidation of water, results in the breaking of hydrogen bonding. Therefore, the system changes from relative order to less order. As a result, the entropy of the system increases, whereas the enthalpy of the system decreases. The overall result leads to the Gibbs free energy to be negative and the process results in the spontaneous formation of peroxide in the presence of oxygen.

Herein, we report on results of a comprehensive dispersion-corrected density-functional method for transition state calculations of the catalytic properties of the negative ions of atomic Au and Pd in the oxidation of water to peroxide, along with the extracted energy barriers. Figure 1 shows the energy of the reactants, transition state, and products in the absence of the atomic Au⁻ or Pd⁻; the energies were calculated in eV. To find the energy barrier, the energy of the product was subtracted from that of the transition state, and the energy was converted to kcal/mole. For this case, the energy barrier was calculated to be 1.75 eV (40.05 kcal/mol).

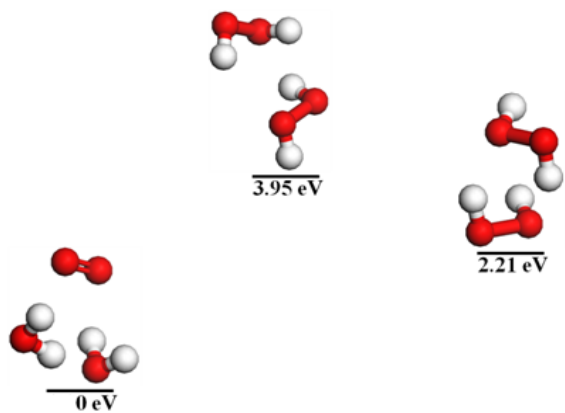

Fig. 1: Transition state of the oxidation of water to peroxide in the absence of a catalyst. The red and white spheres represent O and H, respectively.

Similar calculations were performed in the presence of the Au⁻ and Pd⁻ nanoparticle catalysts as shown in Figure 2 and Figure 3, respectively. The energy barriers were found to be 0.54 eV (12.42 kcal/mol) for Au⁻ and 0.18 eV (4.12 kcal/mol) for Pd⁻ nanoparticle catalysts. The energy barrier in the presence of Au⁻ is almost 3.2 times lower than that of the reaction without the nanocatalyst, and the energy barrier in the presence of Pd⁻ is nearly 3 times lower than that in the presence of Au⁻. It can then be concluded that although the Au⁻ anion nanoparticle speeds up the oxidation of water to peroxide, Pd⁻ possesses a higher catalytic activity than Au⁻ when catalyzing $H_2O_2$ consistent with the recent experimental findings [9, 16].

We further elaborate on the slow oxidation of water to peroxide in the absence of a nanoparticle catalyst. Upon the addition of Au⁻ or Pd⁻ an anion complex is formed with a transition state Au⁻$(H_2O)_2$ or Pd⁻$(H_2O)_2$, respectively. Then the large EAs of the nanoparticles play the important roles in the breaking of the complex to form the Au⁻ or Pd⁻ and $H_2O_2$ products. Although the first and second R-T minima in the TCSs for both Au and Pd are qualitatively the same [24], there is little understanding why Pd⁻ has a higher catalytic activity than Au⁻.

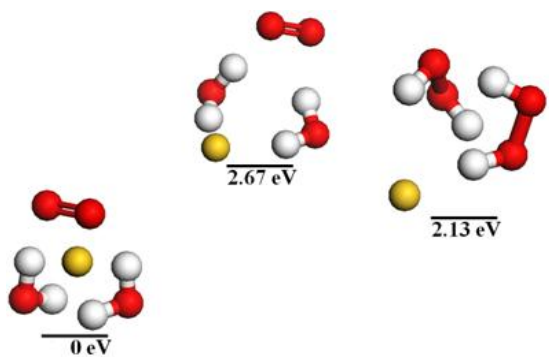

Fig. 2: Transition state of the oxidation of water catalyzed by atomic $Au^-$ to peroxide.   The gold, red and white spheres represent $Au^-$, O, and H, respectively.

A catalyst speeds up a chemical reaction leaving the initial and final states of the reaction intact. However, as shown in Fgures 1, 2, and 3, the energy of the product in the absence and presence of the atomic $Au^-$ and atomic $Pd^-$ catalysts are different; they are found to be 2.21, 2.13, and 2.04 eV, respectively. This can be explained through the effect of the atomic particle catalysts on the strength of hydrogen bonding of the water molecule. Hydrogen bonds arise in water where each partially positively charged hydrogen atom is covalently attached to a partially negatively charged oxygen from a water molecule with bond energy of about 117 kcal/mol and is also attracted, but much more weakly, to a neighboring partially negatively charged oxygen atom from another water molecule.

In liquid water, the energy of attraction between water molecules (hydrogen bond enthalpy) is optimal, about 5.5 kcal/mol.   Breaking one bond generally weakens those around whereas creating one bond generally strengthens those around. This encourages cluster formation where all water molecules are linked together by three or four strong hydrogen bonds. However, in the presence of an atomic particle catalyst such as the atomic $Au^-$, the complex $Au^-(H_2O)_2$ is formed in the transition state. In this anion complex, the two water molecules are attached to the atomic $Au^-$ ion.   At least 30% of the hydrogen bonding strength is broken, which is almost equivalent to 1.64 kcal/mol energy of attraction. The replacement of $Au^-$ by $Pd^-$ breaks the hydrogen bonding to a different extent, with about 50% (2.86 kcal/mol) energy of attraction. This difference can be explained from the electron configurations of both the $Au^-$ and $Pd^-$ ions as follows.

The atomic $Au^-$ ion has a $6s^2 4f^{14} 5d^{10}$ configuration which is of the same type as that of the Xe stable noble gas atom. As a result, during the formation of the $Au^-(H_2O)_2$ complex in the transition state there is a competition of bond making between the atomic $Au^-$ higher energy orbital such as the 6p or 5f and hydrogen for the oxygen's lone pair electrons from a neighboring water molecule. As already pointed out the $Au^-O$ interaction is weaker than the hydrogen bonding [12]. However, in the case of the atomic $Au^-$, which has an electron configuration of $5s^2 4d^9$, the $Pd^-O$ interaction competes strongly with the hydrogen bonding for the lone pair of oxygen from the neighboring water molecule to attain a complete filling of the orbitals. Consequently, the atomic $Pd^-$ ion breaks the hydrogen bonding more strongly than the $Au^-$ atomic particle.   The differences in the energies of the products indicated by our results are attributed directly to this phenomenon.

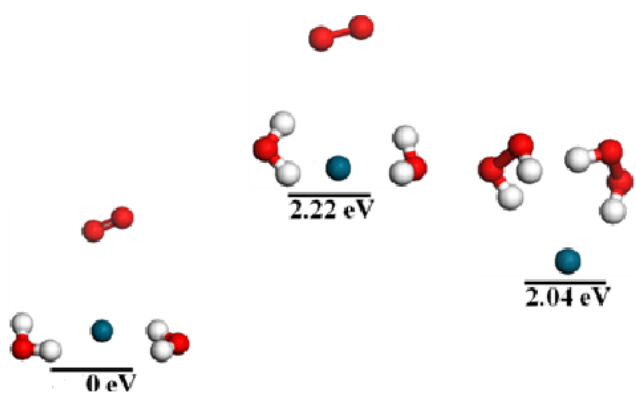

Fig. 3: Transition state of the oxidation of water catalyzed by atomic $Pd^-$ to peroxide. The blue-green, red and white spheres represent $Pd^-$, O, and H, respectively.

Hydrogen bonding has a direct impact on the thermodynamic properties of the reaction. A change in the Gibbs free energy, $G$ ($\Delta G = \Delta H - T\Delta S$) can be used to describe the reaction mechanism upon breaking of the hydrogen bonding. Consider the oxidation of water into peroxide in the absence of the atomic catalyst as the reference point. When the atomic $Au^-$ catalyst is added there will be a breaking of hydrogen bonding, resulting in an increase in the entropy and a decrease in the enthalpy. This leaves the $\Delta G$ more negative compared to the case of the absence of a catalyst, resulting in thermodynamically favorable formation of peroxide. The same explanation can be given for the addition of the atomic $Pd^-$ ion instead of $Au^-$. The percentage of breaking hydrogen bonding is greater for atomic $Pd^-$ compared to atomic $Au^-$. This favors the atomic $Pd^-$ to have a higher catalytic activity than atomic $Au^-$ when catalyzing $H_2O_2$.

To understand the catalytic activities of atomic $Au^-$ and atomic $Pd^-$ in the oxidation of water to peroxide dispersion-corrected density-functional theory calculations for transition states have been performed. The results indicate that although atomic $Pd^-$ speeds up the reaction by lowering the activation energy by about three times that of atomic $Au^-$, both atomic negative ion catalysts exhibit an excellent catalytic effect in the formation of peroxide from water. The fundamental explanation for the excellent catalytic activity of the Au and Pd nanoparticles in the experiments [9, 16] can now be understood from the fundamental atomic physics theory and theoretical chemistry considerations. Crucially, in the presence of an atomic negative ion catalyst such as the atomic $Au^-$ ion, formed through the slow electron-Au collisions, the anionic complex $Au^-(H_2O)_2$ is formed in the transition state. In this anion complex the atomic $Au^-$ ion breaks the hydrogen bonding strength in the two water molecules permitting the formation of $H_2O_2$ in the presence of oxygen.

## 4. Conclusions

The intriguing and puzzling problem of nanocatalysis of $H_2O_2$ from $H_2O$ using the Au and Pd nanoparticles is now understood, for the first time, at the most fundamental level through the combined recent theoretical atomic physics [24,25] and the present theoretical chemistry perspectives. The analyses are general and demonstrate the fundamental mechanism of nanocatalysis through negative ion resonances. The present work certainly ushers in new frontiers of efficient design and synthesis atom by atom of novel functional compounds and catalysts for various chemical reactions, impacting many industries.

In conclusion, we surmise that whenever water is the medium, the atomic $Au^-$ ion will almost always catalyze a reaction, but the atomic $Pd^-$ ion acts as the better and cheaper catalyst than the $Au^-$ ion. We are currently investigating other atoms for use as nanocatalysts for various chemical reactions. A few candidates, such as Au, Ag, Pt, Ru, Y, Tl, At, and a combination of two or more atoms such as Au-Pd have already been identified as possible candidates [17].

**Acknowledgements**

Research was supported by Army Research Office (Grant W911NF-11-1-0194); National Science Foundation (Grant DMR-0934142); U.S. DOE Office of Science; AFOSR (Grants FA9550-10-1-0254 and FA9550-09-1-0672) and the CAU CFNM, NSF-CREST Program.